\documentclass{article}
\usepackage{natbib}
\usepackage{graphicx}
\usepackage{xcolor}
\usepackage{euscript}
\usepackage{amsmath}
\usepackage{epstopdf, epsfig}
\usepackage{amssymb}
\usepackage{graphics}

\title{Frequency downshift in a viscous fluid}

\author{J.D.~Carter \\ email: carterj1@seattleu.edu
\and A.~Govan \\email: govana@seattleu.edu} 
\begin{document}

\maketitle

\section{Abstract}
In this paper, we derive a viscous generalization of the \cite{dysthe}
system from the weakly viscous generalization of the Euler equations
introduced by \cite{DDZ}.  This ``viscous Dysthe'' system models the
evolution of a weakly viscous, nearly monochromatic wave train on deep
water.  It contains a term which provides a mechanism for frequency
downshifting in the absence of wind and wave breaking.  The equation
does not preserve the spectral mean.  Numerical simulations
demonstrate that the spectral mean typically decreases and that the
spectral peak decreases for certain initial conditions.  The linear
stability analysis of the plane-wave solutions of the viscous Dysthe
system demonstrates that waves with wave numbers closer to zero decay
more slowly than waves with wave numbers further from zero.
Comparisons between experimental data and numerical simulations of the
NLS, dissipative NLS, Dysthe, and viscous Dysthe systems establish
that the viscous Dysthe system accurately models data from experiments
in which frequency downshifting was observed {\emph{and}} experiments
in which frequency downshift was not observed.

\section{Introduction}

In the late 1970s, \cite{Lakeplus} and \cite{LY} conducted physical
experiments that investigated the evolution of nonlinear wave trains
on deep water.  They found that the growth of the Benjamin-Feir
instability is followed by a shift in the
spectral peak to a frequency closer to zero.  Subsequent experiments,
including those in \cite{Suetal} and \cite{Melville}, demonstrated
that the amplitude of the lower sideband grows and eventually
overtakes that of the carrier wave.  These later experimental studies
focused on waves with larger steepness and involved wave breaking.
More recently, \cite{sh} conducted similar experiments without wave
breaking or wind.  They found that frequency downshifting (FD) is not
observed (in their tank) if the waves have ``small or moderate''
amplitudes and that FD is observed if the amplitude of the carrier
wave is ``large'' or if the sideband perturbations are ``large
enough.''  They also found that if FD occurs then (i) the spectral
mean (defined below) decreases monotonically in time and (ii) FD
occurs in the higher harmonics before it occurs in the fundamental.
The goal of the current work is to provide a mathematical
justification for FD that does not rely on wind or wave breaking.
Readers interested in wind and wave-breaking justifications for FD are
referred to~\cite{TD1990,Hara,Kato,IslasSchober}.

There are two quantities that are commonly used to quantify frequency
downshifting: the spectral peak and the spectral mean.  The spectral
peak, $k_p$, is defined to be the wave number corresponding to the
Fourier mode with largest magnitude.  The spectral mean, $k_m$, is
defined by
\begin{equation}
k_m=\frac{\mathcal{P}}{\mathcal{M}},
\label{kc}
\end{equation}
where $\mathcal{M}$, the ``mass,'' and $\mathcal{P}$, the ``linear
momentum,'' are defined by
\begin{subequations}
\begin{equation}
\mathcal{M}=\frac{1}{L}\int_0^L|\eta|^2d\xi,
\end{equation}
\begin{equation}
\mathcal{P}=\frac{i}{2L}\int_0^L\big{(}\eta\eta_\xi^*-\eta_\xi \eta^*\big{)}d\xi,
\end{equation}
\label{MP}
\end{subequations}
where $\eta$ and $L$ represent the free surface displacement and
period of the solution/wave train respectively.  We note that while  
some studies use $\mathcal{P}$ by itself as a measure of FD, we focus
on $k_p$ and $k_m$.

\cite{Zak1968} derived the cubic nonlinear Schr\"odinger equation
(NLS) from the Euler equations as a model for the evolution of the
envelope of a nearly monochromatic wave group.  NLS preserves the
spectral mean, so it cannot be used to model FD.  \cite{dysthe}
carried out the NLS perturbation analysis one order higher to obtain
what is now known as the Dysthe system.  \cite{lomei} numerically
solved NLS and the Dysthe system and established that the Dysthe system
more accurately predicts the evolution of mildly sloped, narrow-banded,
weakly nonlinear waves over longer time periods than does NLS.  They
also found that dissipative generalizations of the Dysthe system are
required to model waves of moderate steepness over long distances.
Finally, their numerical studies showed that the Dysthe system did not
lead to a permanent FD even though the Dysthe system does not preserve
the spectral mean.  \cite{sh} established that the dissipative NLS equation
accurately modeled the evolution of waves trains in which no FD
occurred and that it cannot model FD because it preserves the spectral
mean.  The dissipative NLS equation was used as an ad-hoc model
without formal justification until \cite{DDZ} derived it from a weakly
viscous generalization of the Euler equations.  The first step in the
current work is to carry out the Dysthe perturbation analysis starting
from the \cite{DDZ} weakly viscous generalization of the Euler
equations in order to derive a new system which we call the viscous
Dysthe system.

The paper is organized as follows.  Section \ref{derivation} contains
the derivation of the viscous Dysthe system.  Section \ref{properties}
contains a summary of the main properties of this new equation.
Section \ref{experiments} contains comparisons of viscous Dysthe
predictions  with experimental data from two experiments, one of which
exhibiting FD.

\section{Derivation of the viscous Dysthe equation}
\label{derivation}

\cite{DDZ} introduced the following system for an
infinitely-deep weakly viscous fluid
\begin{subequations}
\begin{equation}
\phi_{xx}+\phi_{yy}+\phi_{zz}=0, \hspace*{1cm}\mbox{for } -\infty<z<\eta,
\end{equation}
\begin{equation}
\phi_t+\frac{1}{2}\big{|}\nabla\phi\big{|}^2+g\eta=-2\bar{\nu}\phi_{zz},\hspace*{1cm}\mbox{at
} z=\eta,
\end{equation}
\begin{equation}
\eta_t+\eta_x\phi_x+\eta_y\phi_y=\phi_z+2\bar{\nu}\Delta\eta,\hspace*{1cm}\mbox{at } z=\eta,
\label{Diasc}
\end{equation}
\begin{equation}
|\nabla\phi|\rightarrow0, \hspace*{1cm}\mbox{as } z\rightarrow-\infty.
\end{equation}
\label{DDZ}
\end{subequations}
Here $\phi=\phi(x,y,z,t)$ is the velocity potential of the fluid,
$\eta=\eta(x,y,t) $ is the free-surface displacement, $g$ is the
acceleration due to gravity, and $\bar{\nu}>0$ is the kinematic
viscosity of the fluid.  This model assumes that gravity and
viscosity are the only external forces acting on the fluid.  The Euler
equations are obtained from (\ref{DDZ}) by setting $\bar{\nu}=0$.

Substituting 
\begin{subequations}
\begin{equation}
\phi=\phi_0\mbox{e}^{ik_0x+k_0z-i\lambda_0 t},
\end{equation}
\begin{equation}
\eta=\eta_0\mbox{e}^{ik_0x-i\lambda_0 t},
\end{equation}
\end{subequations}
into the linearized version of (\ref{DDZ}) gives its linear
``dispersion'' relation 
\begin{equation}
\lambda_0=\pm\sqrt{gk_0}-2ik_0^2\bar{\nu}.
\label{DDZdispersion}
\end{equation}
The $\pm\sqrt{gk_0}$ term represents dispersion while the
$-2ik_0^2\bar{\nu}$ term represents (wave-number dependent)
dissipation.  This establishes that, according to this model,
small-amplitude waves with large wave number will decay faster than
small-amplitude waves with small wave number.

Following the work of \cite{dysthe}, assume
\begin{subequations}
\begin{equation}
\eta(x,y,t)=\epsilon^3\bar{\eta}+\epsilon B\mbox{e}^{ik_0x-i\omega_0
  t}+\epsilon^2B_2\mbox{e}^{2(ik_0x-i\omega_0t)}
%+\epsilon^3B_3\mbox{e}^{3(ik_0x-i\omega_0 t)}
+\dots+c.c.,
\label{eta}
\end{equation}
\begin{equation}
\phi(x,y,z,t)=\epsilon^2\bar{\phi}+\epsilon A_1\mbox{e}^{k_0z+ik_0x-i\omega_0
  t}+\epsilon^2 A_2\mbox{e}^{2(k_0z+ik_0x-i\omega_0t)}
%\epsilon^3 A_3\mbox{e}^{3(k_0z+ik_0x-i\omega_0t)}
+\dots+c.c.,
\end{equation}
\label{ansatz}
\end{subequations}
where $k_0$ is a positive constant and $\epsilon=a_0k_0\ll 1$ is a small
dimensionless parameter.  Here $a_0$ represents a typical amplitude,
$k_0$ represents the wave number of the carrier wave, and $c.c.$ stands
for complex conjugate.  The $A$'s and $\bar{\phi}$ depend on the slow
variables $X=\epsilon x$, $Y=\epsilon Y$, $Z=\epsilon x$, and
$T=\epsilon t$, while the $B$'s and $\bar{\eta}$ depend on $X$, $Y$ and
$T$.  Although it is done without loss of generality, assuming $k_0>0$
is significant because it determines the form of the $z$ dependence in
$\phi$.  Next, assume
\begin{subequations}
\begin{equation}
A_j=A_{j0}+\epsilon A_{j1}+\epsilon^2 A_{j2}+\epsilon^3
A_{j3}+\dots,\hspace*{1cm}\mbox{for }j=1,2,3,\dots,
\end{equation}
\begin{equation}
B_j=B_{j0}+\epsilon B_{j1}+\epsilon^2 B_{j2}+\epsilon^3
B_{j3}+\dots,\hspace*{1cm}\mbox{for }j=2,3,4,\dots,
\end{equation}
\begin{equation}
\bar{\eta}=\bar{\eta}_{0}+\epsilon \bar{\eta}_{1}+\epsilon^2 \bar{\eta}_{2}+\dots,
\end{equation}
\begin{equation}
\bar{\phi}=\bar{\phi}_{0}+\epsilon \bar{\phi}_{1}+\epsilon^2 \bar{\phi}_{2}+\dots.
\end{equation}
\label{ansatz2}
\end{subequations}
Following the work of \cite{DDZ}, assume that viscosity effects are
small by assuming $\bar{\nu}=\epsilon^2\nu$.  Substituting
(\ref{ansatz})-(\ref{ansatz2}) into (\ref{DDZ}) and carrying out the
perturbation analysis through $\mathcal{O}(\epsilon^4)$ gives the
deep-water dispersion relationship,
\begin{equation}
\omega_0^2=gk_0,
\label{disprel}
\end{equation} 
the system that defines $B$ and $\bar{\phi}_0$,
\begin{subequations}
\begin{align}
\begin{split}
\hspace*{-3cm}2i\omega_0\big{(}B_T&+\frac{g}{2\omega_0}B_X\big{)}+\epsilon\Big{(}-\frac{g}{4k_0}
B_{XX}+\frac{g}{2k_0}B_{YY}-4gk_0^3|B|^2B+4ik_0^2\nu
  B\Big{)}\\
&+\epsilon^2\Big{(}-i\frac{g}{8k_0^2}B_{XXX}+i\frac{3g}{4k_0^2}B_{XYY}+2igk_0^2B^2B^*_X\\
&\hspace*{0.5cm}+12igk_0^2|B|^2B_x-2k_0\omega_0
B\bar\phi_{0X}+8k_0\omega_0\nu B_X\Big{)}=0,\hspace*{0.2cm}{\text{at
    }}Z=0,
\label{vBstationary}
\end{split}
\end{align}
\begin{equation}
\bar{\phi}_{0Z}=2\omega_0\Big{(}|B|^2\Big{)}_X,\hspace*{1cm}{\text{at
    }}Z=0,
\end{equation}
\begin{equation}
\bar{\phi}_{0XX}+\bar{\phi}_{0ZZ}=0,\hspace*{1cm}{\text{for}} -\infty<Z<0,
\end{equation}
\begin{equation}
\bar{\phi}_{0Z}\rightarrow 0,\hspace*{1cm}{\text{as }} Z\rightarrow-\infty,
\end{equation}
\label{vBsystem}
\end{subequations}
and 
\begin{subequations}
\begin{equation}
B_{2}=k_0B^2+\mathcal{O}(\epsilon),
\label{B2}
\end{equation}
\begin{equation}
B_{3}=\frac{3}{2}k_0^2B^3+\mathcal{O}(\epsilon).
\label{B3}
\end{equation}
\label{B2B3}
\end{subequations}
Equations (\ref{B2}) and (\ref{B3}) define the leading-order
contributions to the amplitudes of the second and third harmonics of
the carrier wave respectively.  Following a similar procedure, but
solving for the leading-order term of the velocity potential instead
of the leading-order term of the surface displacement, gives
\begin{subequations}
\begin{align}
\begin{split}
\hspace*{-3cm}2i\omega_0\big{(}A_T&+\frac{g}{2\omega_0}A_X\big{)}+\epsilon\Big{(}-\frac{g}{4k_0}
A_{XX}+\frac{g}{2k_0}A_{YY}-4k_0^4|A|^2A+4ik_0^2\nu
  A\Big{)}\\
&+\epsilon^2\Big{(}-i\frac{g}{8k_0^2}A_{XXX}+i\frac{3g}{4k_0^2}A_{XYY}-2ik_0^3A^2A^*_X\\
&\hspace*{0.5cm}+12ik_0^3|A|^2A_x-2k_0\omega_0
A\bar\phi_{0X}+8
  k_0\omega_0\nu A_X\Big{)}=0,\hspace*{0.2cm}{\text{at
    }}Z=0,
\label{vAstationary}
\end{split}
\end{align}
\begin{equation}
\bar{\phi}_{0Z}=\frac{2k_0^2}{\omega_0}\Big{(}|A|^2\Big{)}_X,\hspace*{1cm}{\text{at
    }}Z=0,
\end{equation}
\begin{equation}
\bar{\phi}_{0XX}+\bar{\phi}_{0ZZ}=0,\hspace*{1cm}\hspace*{1cm}{\text{for}} -\infty<Z<0,
\end{equation}
\begin{equation}
\bar{\phi}_{0Z}\rightarrow 0,\hspace*{1cm}{\text{as }} Z\rightarrow-\infty.
\end{equation}
\label{vAsystem}
\end{subequations}
The systems in (\ref{vBsystem}) and (\ref{vAsystem}) are related by
\begin{subequations}
\begin{equation}
B=\frac{ik_0}{\omega_0}A+\epsilon\frac{1}{2\omega_0}A_X+\epsilon^2\Big{(}\frac{i}{8k_0\omega_0}A_{XX}-\frac{i}{4k_0\omega_0}A_{YY}+\frac{ik_0^4}{2g\omega_0}|A|^2A\Big{)}+\mathcal{O}(\epsilon^3),
\label{BA}
\end{equation}
\begin{equation}
A=-\frac{i\omega_0}{k_0}B+\epsilon\frac{\omega_0}{2k_0^2}B_X+\epsilon^2\Big{(}\frac{3i\omega_0}{8k_0^3}B_{XX}-\frac{i\omega_0}{4k_0^3}B_{YY}+\frac{ik_0\omega_0}{2}|B|^2B\Big{)}+\mathcal{O}(\epsilon^3).
\label{AB}
\end{equation}
\label{ABBA}
\end{subequations}

We now change variables in order to enter into a coordinate frame
moving with the linear group velocity, $c_g=\frac{\omega_0}{2k_0}$.
For simplicity, we also assume that there is no $Y$ dependence.  Let
\begin{subequations}
\begin{equation}
B(X,Y,T)=\tilde{B}(\xi,\chi),
\end{equation}
\begin{equation}
A(X,Y,Z,T)=\frac{\omega_0}{k_0}\tilde{A}(\xi,\chi,\zeta),
\end{equation}
\begin{equation}
\bar\phi_0(X,Y,Z,T)=4\omega_0\tilde{\Phi}(\xi,\chi,\zeta),
\end{equation}
\begin{equation}
\nu=\frac{\omega_0}{4k_0^2}\delta,
\end{equation}
\begin{equation}
\chi=\epsilon k_0X,
\end{equation}
\begin{equation}
\xi=2k_0X-\omega_0T,
\label{xiCOV}
\end{equation}
\begin{equation}
\zeta=k_0Z.
\end{equation}
\label{COV}
\end{subequations}
This change of variables leads to (where tildes have been dropped for
convenience)
\begin{subequations}
\begin{equation}
iB_\chi-B_{\xi\xi}-4k_0^2|B|^2B+i\delta
B+\epsilon\big{(}8ik_0^2B^2B^*_\xi+32ik_0^2|B|^2B_\xi-16k_0^2
B\Phi_\xi+5\delta B_\xi\big{)}=0,\hspace*{0.1cm}{\text{at
    }}\zeta=0,
\label{vBalone}
\end{equation}
\begin{equation}
\Phi_\zeta=\Big{(}|B|^2\Big{)}_\xi,\hspace*{1cm}{\text{at
    }}\zeta=0,
\end{equation}
\begin{equation}
4\Phi_{\xi\xi}+\Phi_{\zeta\zeta}=0,\hspace*{1cm}\hspace*{1cm}{\text{for}} -\infty<\zeta<0,
\end{equation}
\begin{equation}
\Phi_\zeta\rightarrow 0,\hspace*{1cm}{\text{as }} \zeta\rightarrow-\infty,
\end{equation}
\label{vB}
\end{subequations}
and
\begin{subequations}
\begin{equation}
iA_\chi-A_{\xi\xi}-4k_0^2|A|^2A+i\delta
A+\epsilon\big{(}32ik_0^2|A|^2A_\xi-16k_0^2A\Phi_\xi+5\delta A_\xi\big{)}=0,\hspace*{0.1cm}{\text{at
    }}\zeta=0,
\label{vAalone}
\end{equation}
\begin{equation}
\Phi_\zeta=\big{(}|A|^2\big{)}_\xi,\hspace*{1cm}{\text{at
    }}\zeta=0,
\end{equation}
\begin{equation}
4\Phi_{\xi\xi}+\Phi_{\zeta\zeta}=0,\hspace*{1cm}\hspace*{1cm}{\text{for}} -\infty<\zeta<0,
\end{equation}
\begin{equation}
\Phi_\zeta\rightarrow 0,\hspace*{1cm}{\text{as }} \zeta\rightarrow-\infty.
\end{equation}
\label{vA}
\end{subequations}
We call the systems given in equations (\ref{vB}) and (\ref{vA}) the
viscous Dysthe system for the surface displacement (the vB system) and
the viscous Dysthe system for the velocity potential (the vA system)
respectively.  These equations, which describe the evolution of a
narrow-banded, weakly nonlinear, weakly viscous fluid and (as shown
below) accurately predict FD, are the main results of this paper.
NLS is obtained from equation (\ref{vBalone}) or (\ref{vAalone}) by
setting $\delta=\epsilon=0$.  The classical Dysthe systems for the
surface displacement and velocity potential are obtained from
(\ref{vB}) and (\ref{vA}) by setting $\delta=0$.

It is important to note that there is only one free parameter in the
viscous Dysthe systems.  This parameter, $\delta$, is empirically
determined by fitting the decay of $\mathcal{M}$.  This will
establish (see below) that changes in $\mathcal{M}$, $\mathcal{P}$,
and $k_m$ are all determined by $\delta$.  This is qualitatively
different from other FD models that have two (or more) free parameters
that can be used to fit the decay of both $\mathcal{M}$ and
$\mathcal{P}$.

\section{Properties of the viscous Dysthe system}
\label{properties}

\subsection{Mass and linear momentum}

\cite{TD} showed that the Dysthe system for the
velocity potential (the vA system with $\delta=0$) preserves both
$\mathcal{M}$ and $\mathcal{P}$ in $\chi$ while the Dysthe system for
the surface displacement (the vB  system with $\delta=0$) preserves
$\mathcal{M}$, but not $\mathcal{P}$.  Moreover, they established that
the sign of $\mathcal{P}_\chi$ in the vB system with $\delta=0$ is not
definite and therefore the Dysthe system can predict frequency
upshifting.  \cite{lomei} conducted numerical studies of the
vB system with $\delta=0$ and did not observe a permanent downshift in
any of their simulations.  These results suggest that the $B^2B^*_\xi$
term (the only term that differentiates the vA and vB systems when
$\delta=0$) is not the mechanism for FD.

We note that the vA system with $\delta=0$ is Hamiltonian while the vB
system with $\delta=0$ is likely not
Hamiltonian~\citep[see][]{thesis,shh}. \cite{GT} derived a Hamiltonian
generalization of the Dysthe system for the  surface displacement.
Although we do not study this equation here, we expect that the
results shown below will generalize to it because the
$\mathcal{O}(\epsilon)$ viscosity term, $5\epsilon\delta B_\xi$,
provides the mechanism for FD.

The $\chi$ dependences of $\mathcal{M}$ and $\mathcal{P}$ in the vB
system are given by 
\begin{subequations}
\begin{equation}
\mathcal{M}_\chi=-2\delta\mathcal{M}-10\epsilon\delta\mathcal{P},
\label{dMA}
\end{equation}
\begin{equation}
\mathcal{P}_\chi=-2\delta\mathcal{P}-10\epsilon\delta\mathcal{Q}-16\epsilon
k_0^2\mathcal{R},
\label{dPA}
\end{equation}
\label{dCQsA}
\end{subequations}
where
\begin{subequations}
\begin{equation}
\mathcal{Q}=\frac{1}{L}\int_0^L |B_\xi|^2d\xi,
\label{Q}
\end{equation}
\begin{equation}
\mathcal{R}=\operatorname{Re}\Big{(}\frac{1}{L}\int_0^L|B|^2BB^*_{\xi\xi}d\xi\Big{)}.
\label{R}
\end{equation}
\end{subequations}
Further
\begin{equation}
\big{(}k_m\big{)}_\chi=\Big{(}\frac{\mathcal{P}}{\mathcal{M}}\Big{)}_\chi=-\frac{10\epsilon\delta}{\mathcal{M}^2}\Big{(}\mathcal{M}\mathcal{Q}-\mathcal{P}^2\Big{)}-16\epsilon
k_0^2\frac{\mathcal{R}}{\mathcal{M}}.
\label{dPM}
\end{equation}
The Cauchy-Schwarz inequality establishes that
$(\mathcal{M}\mathcal{Q}-\mathcal{P}^2)>0$.  This suggests that the vB
system will exhibit FD in the spectral mean sense.  However, the sign of
$\mathcal{R}$ can vary, so FD is not necessarily guaranteed.  Note
that FD is guaranteed in the vA system because the $\mathcal{R}$
term is not involved.  FD does not occur in either the vA or the vB
system if either i) $\delta=0$ (i.e.~$\nu=0$) or ii) $\epsilon=0$ (not
physically interesting).  Finally, equation (\ref{dPM}) establishes
that FD is a higher-order effect, corroborating the \cite{sh} result that the dissipative NLS equation preserves
the spectral mean.

\subsection{Plane-Wave Solutions}

Consider solutions of the vB system with the following form
\begin{equation}
B(\xi,\chi)=B_0\exp\big{(}il\xi+w_r(\chi)+iw_i(\chi)+i\rho\big{)},
\label{SWSoln}
\end{equation}
where $B_0$, $l$, and $\rho$ are real constants and $w_r$ and
$w_i$ are real-valued functions.  This gives
\begin{subequations}
\begin{equation}
\Phi(\xi,\chi,\zeta)=0,
\end{equation}
\begin{equation}
w_r^{\prime}+\delta(1+5\epsilon l)=0,
\end{equation}
\begin{equation}
w_i^{\prime}-l^2+4B_0^2k_0^2(1+6\epsilon l)\mbox{e}^{2w_r}=0.
\end{equation}
\label{SWODEs}
\end{subequations}
The parameter $\rho$ does not play a role in the ODEs that define
$\omega_r$ and $\omega_i$ because of the gauge invariance of the vB
system.  This allows all complex $B_0$ to be considered by
assuming, without loss of generality, that $B_0$ is a positive
constant.  The solution of this system is
\begin{subequations}
\begin{equation}
w_r(\chi)=-\delta(1+5\epsilon l)\chi,
\label{omegar}
\end{equation}
\begin{equation}
w_i(\chi)=l^2\chi+\frac{2A_0^2k_0^2(1+6\epsilon l)}{\delta(1+5\epsilon
  l)}\Big{(}\mbox{e}^{-2\delta(1+5\epsilon l)\chi}-1\Big{)}.
\end{equation}
\label{wrwi}
\end{subequations}
The constants of integration were chosen so that these (viscous)
solutions limit to the solutions of the (nonviscous) Dysthe system in
the $\delta\rightarrow0$ limit.  Note that (\ref{omegar}) establishes
that if $\delta>0$, then to leading order all plane-wave solutions of the
vB system decay to zero as $\chi\rightarrow\infty$.  Just as in NLS
and the Dysthe system, choosing $l\ne0$ corresponds to shifting the
wave number of the carrier wave.  The function $\omega_r$ depends on
$l$ in such a way that it takes into account the correct (up to the order
of the equation) rate of dissipation for the shifted carrier wave (see
equations (\ref{DDZdispersion}) and (\ref{COV})).  Therefore, we
assume $l=0$ for the remainder of the paper.

\subsection{Linear stability analysis}
In order to study the stability of solutions of the form given in
equation (\ref{SWSoln}), we consider perturbed solutions of the form
\begin{subequations}
\begin{equation}
B_{\text{pert}}(\xi,\chi)=\big{(}B_0+\mu u(\xi,\chi)+i\mu
v(\xi,\chi)\big{)}\exp\big{(}w_r(\chi)+iw_i(\chi)\big{)},
\label{B0p}
\end{equation}
\begin{equation}
\Phi_{\text{pert}}(\xi,\chi,\zeta)=0+\mu p(\xi,\chi,\zeta),
\end{equation}
\label{pertform}
\end{subequations}
where $\mu$ is a small real parameter; $u$, $v$, and $p$ are
real-valued functions; and $w_r(t)$ and $w_i(t)$ are given in equation
(\ref{wrwi}).  Note that the decay due to $w_r(t)$ has been factored
out in equation (\ref{B0p}).  This allows the evolution of the
perturbation, $u+iv$, to be directly  compared to the constant $B_0$.
If $u$ and/or $v$ grow without bound in $\chi$, we say that the
solutions is linearly unstable (even though the magnitude of the
entire perturbed solution may decay to zero).

Substituting (\ref{pertform}) into (\ref{vB}), linearizing in $\mu$,
and separating into real and imaginary parts leads to a system of
coupled PDEs that have constant coefficients in $\xi$ and $\zeta$.  Without
loss of generality, assume
\begin{subequations}
\begin{equation}
u(\xi,\chi)=U(\chi)\mbox{e}^{iq\xi}+U^*(\chi)\mbox{e}^{-iq\xi},
\end{equation}
\begin{equation}
v(\xi,\chi)=V(\chi)\mbox{e}^{iq\xi}+V^*(\chi)\mbox{e}^{-iq\xi},
\end{equation}
\begin{equation}
p(\xi,\chi,\zeta)=\mbox{sign}(q)B_0\big{(}iU(\chi)\mbox{e}^{iq\xi}-iU^*(\chi)\mbox{e}^{-iq\xi}\big{)}\mbox{e}^{2|q|\zeta}\mbox{e}^{-2\delta\chi},
\end{equation}
\label{pertsolns}
\end{subequations}
where $q$ is a real constant, and $U$ and $V$ are complex-valued
functions.  The form for $p$ was selected because it satisfies
Laplace's equation and the bottom boundary condition.  Substituting
(\ref{pertsolns}) into the PDEs gives 
\begin{equation}
\begin{pmatrix} U \\ V\end{pmatrix}^{\prime}=\Big{(}\mathbf{A_1}+\mbox{e}^{-2w_r(\chi)}\mathbf{A_2}\Big{)}\begin{pmatrix} U \\ V\end{pmatrix},
\label{ODEs}
\end{equation}
where prime means derivative with respect to $\chi$ and $\mathbf{A_1}$
and $\mathbf{A_2}$ are the constant matrices given by 
\begin{subequations}
\begin{equation}
\mathbf{A_1}=\begin{pmatrix} 0 & -q^2-5i\epsilon\delta q \\
  q^2+5i\epsilon\delta q & 0
   \end{pmatrix},
\end{equation}
\begin{equation}
\mathbf{A_2}=\begin{pmatrix} -40i\epsilon qk_0^2 & 0 \\
  -8k_0^2(1-2\epsilon|q|) & -24i\epsilon qk_0^2
   \end{pmatrix}.
\end{equation}
\end{subequations}
To our knowledge, an exact solution of (\ref{ODEs}) is not known.
However, $\mathbf{A}_1$ determines the large-$\chi$ behavior of the
solution because the $\chi$ dependent term is integrable on
$\chi\in [0,\infty)$ \citep[see][]{codd}.  The eigenvalues of
$\mathbf{A_1}$ are
\begin{equation}
\lambda_{A_1}=\pm iq^2 \mp 5\epsilon\delta q.
\end{equation}
This establishes that the long-term solution of (\ref{ODEs}) is given
by
\begin{equation}
\begin{pmatrix} U \\ V
\end{pmatrix}
=\begin{pmatrix}
-ic_1 & ic_2 \\
c_1 & c_2
\end{pmatrix}
\begin{pmatrix}
\exp\big{(}(-iq^2+5\epsilon\delta q)\chi\Big{)}\\ \exp\big{(}(iq^2-5\epsilon\delta q)\chi\big{)}
\end{pmatrix},
\end{equation}
where $c_1$ and $c_2$ are the constants of integration.  Substituting
this back into equation (\ref{pertsolns}) gives
\begin{equation}
B_{\text{pert}}=\Big{(}B_0+2ic_1\exp\big{(}i(-q\xi+q^2\chi)+5\epsilon\delta
q\chi\big{)}+2ic_2\exp\big{(}i(q\xi+q^2\chi)-5\epsilon\delta
q\chi\big{)}\Big{)}\mbox{e}^{w_r(\chi)+iw_i(\chi)}.
\label{Apert}
\end{equation}

\noindent{\textbf{Observations:}}
\begin{itemize}
\item{If $\delta>0$ and $q\ne0$, then $\mathbf{A}_1$ has an eigenvalue
    with positive real part.  Therefore all plane-wave solutions of
    the vB system are linearly unstable.  The growth rate of the
    instability is $5\epsilon\delta|q|$.  This instability is not a
    Benjamin-Feir-like instability because {\emph{all}} spatially
    dependent perturbations lead to exponential growth.  This
    instability is similar to the ``enhanced Benjamin-Feir''
    instabilities that arise in other dissipative generalizations
    of NLS \citep[see][]{BridDias,CC}.}
\item{If $q$ is positive, then the $c_1$ term in (\ref{Apert}) grows
    exponentially in $\chi$ while the $c_2$ term decays exponentially
    in $\chi$.  If $q$ is negative, then the $c_2$ term grows while
    the $c_1$ term decays.  This intrinsic preference for
    perturbations with negative wave numbers suggests FD.}
\item{In order to make statements about the physical surface
    displacement, one needs to transform back into physical
    coordinates by using equations (\ref{ansatz}), (\ref{ABBA}), and
    (\ref{COV}).  The amplitude of the carrier wave (the mode with
    wave number $k_0>0$) decays exponentially.  The amplitude of the
    upper sideband (the mode with wave number $k_0+|q|$) decays more
    rapidly than the amplitude of the carrier wave.  The amplitude of
    the lower sideband ($k_0-|q|$) decays more slowly than does the
    amplitude of the carrier wave.  This suggests FD.}
\item{In equation (\ref{Apert}) the decay due to $\omega_r$,
    namely $\exp(-\delta\chi)$, has been factored out.  This
    decay is an $\mathcal{O}(1)$ effect, while the growth of the
    instability is an $\mathcal{O}(\epsilon)$ effect.  This means that
    the decay typically dominates the growth.  However, some modes
    decay more slowly than others.  Waves with physical wave numbers
    closer to zero decay more slowly than those with wave numbers 
    further from zero. This suggests FD.}
\item{These results apply to $B$ in (\ref{ansatz}).  Calculations
    for the $B^*$ term in (\ref{ansatz}) establish that the amplitude
    of the mode with wave number $-k_0-|q|$ decays more rapidly than the
  amplitude of the carrier wave while the mode with wave number
  $-k_0+|q|$ decays less rapidly.  This suggests FD.}
\item{The growth rate $5\epsilon\delta|q|$ and equation (\ref{B2B3})
    establish that the instability in the second sideband grows twice
    as fast as the instability in the fundamental.  This suggests that
    FD will be observed in the higher harmonics before it is observed
    in the fundamental (just as was observed by \cite{sh}).}
\end{itemize}

\section{Comparisons with experiments}
\label{experiments}

In this section the validity of the vB system is tested by
comparing its predictions with measurements from two physical
experiments of one-dimensional, nearly monochromatic wave trains on
deep water and predictions from the NLS, dissipative NLS, and Dysthe
equations.  The waves were created by a plunger-type wave maker that
oscillates vertically at one end of a 43-foot long and 10-inch wide
wave channel.  Time series were collected by eleven or twelve
(depending on the experiment) wave gauges located $128+50(n-1)$cm from
the wave maker were  $n=1,\dots,12$.  The tank was long enough that
reflections off the far wall did not play a role.  

The first experiment examines the evolution of a nearly monochromatic
wave train formed by a carrier wave perturbed by sidebands of
``moderate'' amplitude.  The second experiment examines the evolution
of a nearly monochromatic wave train formed by a very similar carrier
wave perturbed by sidebands of ``large'' amplitude.  FD was observed
in the second experiment, but not the first.  For each
experiment, the wave gauges recorded time series lasting 23.4033
seconds.  The experimental initial conditions were comprised of a
carrier wave with a frequency of 3.33Hz ($k_0=0.447475\text{cm}^{-1}$)
and one upper and one lower seeded sideband each separated from the
carrier wave by 0.17092Hz ($\Delta k=0.04273\text{cm}^{-1}$).  Complete
experimental details are found in Section 6 of \cite{sh}.

All equations were solved numerically by assuming periodic boundary
conditions and using split-step pseudospectral methods that allow the
linear parts of the PDEs to be solved exactly in Fourier space.  The
initial conditions for the PDEs were
\begin{equation}
B(\xi,\chi=0)=\sum_{n=-3}^{3}\frac{1}{\epsilon}a_n^*~\mbox{e}^{in\xi/(78\epsilon)}.
\label{ICs}
\end{equation}
This form was selected for the initial conditions because it contains
all of the modes with significant amplitude measured by the first wave
gauge.  The complex conjugates of the amplitudes were used because of
the change of variables in equation (\ref{xiCOV}).  Note that when solving the
NLS or dissipative NLS equation, $\epsilon$ is set to zero in equation
(\ref{vBalone}) in order to determine the correct equation and then a
value of $\epsilon$ from either Table \ref{MATable} or \ref{LATable}
was used in (\ref{ICs}) to obtain the appropriately scaled initial
condition.

\subsection{Moderate-amplitude experiment}

\begin{figure}
\begin{center}
\includegraphics[width=\textwidth]{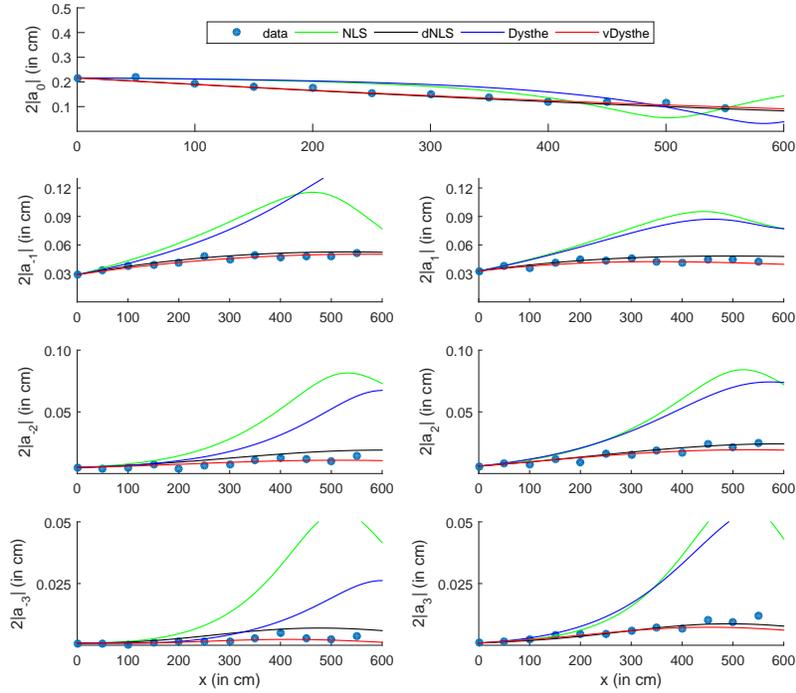}
\caption{Plots of the amplitudes (in cm) of the carrier wave and the
  six nearest sidebands plotted versus the distance (in cm) from
  the first wave gauge for the moderate-amplitude experiment.}
\label{FampsMod}
\end{center}
\end{figure}

Table \ref{MATable} contains the values of the physical parameters for
this experiment.  Figure \ref{FampsMod} contains plots of the Fourier
amplitudes (in centimeters) versus $x$, the distance from the first
wave gauge (in centimeters).  Unlike the plots in~\cite{sh}, the
experimental data here are not scaled.  Although the amplitudes of the
sidebands increase, the amplitude of the carrier wave is dominant
at all measurement sites.  This means that FD in the spectral peak
sense did not occur in this experiment.

\begin{table}
\begin{center}
\begin{tabular}{|c|c|c|}
\hline
parameter & symbol & value\\
\hline
wave number of carrier wave & $k_0$ & 0.447475\\
initial amplitude of carrier wave & $a_0$ & 0.057756+0.091504i\\
initial amplitude of first upper sideband & $a_1$ & 0.0093077-0.013313i\\
initial amplitude of second upper sideband & $a_2$ & -0.0028183-0.0010747i\\
initial amplitude of third upper sideband & $a_3$ & -0.00025316+0.00037694i\\
initial amplitude of first lower sideband & $a_{-1}$ & 0.0035254-0.013840i\\
initial amplitude of second lower sideband & $a_{-2}$ & 0.00033989+0.0024345i\\
initial amplitude of third lower sideband & $a_{-3}$ & 0.00036841+0.00020832i\\
dimensionless parameter & $\epsilon=2k_0|a_0|$ & 0.096840\\
viscosity coefficient & $\delta$ & 0.0011068\\
\hline
\end{tabular}
\caption{Experimental parameters for the moderate-amplitude experiment.}
\label{MATable}
\end{center}
\end{table}

The dissipative theories (viscous Dysthe and dissipative NLS) do a
much better job predicting the evolution of the amplitudes of the
carrier wave and the six nearest sidebands than do
the conservative theories (NLS and Dysthe).  Both conservative models
greatly over predict the growth of the sidebands.  The vB system
provides predictions for all seven amplitudes that are at least as
accurate as, if not more accurate than, the predictions given by the
dissipative NLS equation.  For this experiment, the vA system (results
not shown) provides predictions that are approximately the same as the
vB predictions.

\begin{figure}
\begin{center}
\includegraphics[width=\textwidth]{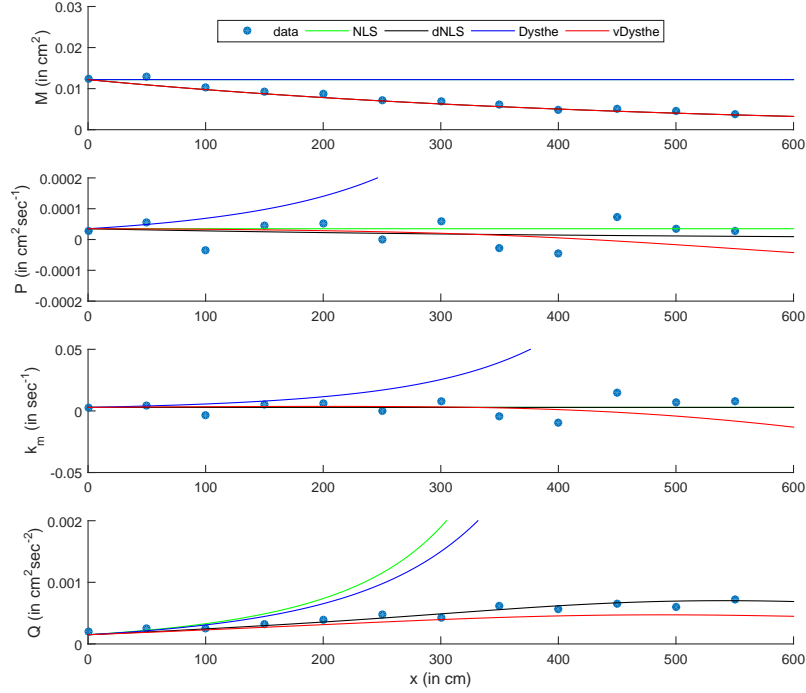}
\caption{Plots of $\mathcal{M}$, $\mathcal{P}$, $k_m$, and
  $\mathcal{Q}$ versus the distance (in cm) from the first wave gauge
  for the moderate-amplitude experiment.}
\label{CQMod}
\end{center}
\end{figure}

Figure \ref{CQMod} shows how the quantities $\mathcal{M}$,
$\mathcal{P}$, $k_m$, and $\mathcal{Q}$ evolved as the wave train
progressed down the tank. The spectral mean, $k_m$, is constant to
within experimental error and therefore FD in the spectral mean sense
did not occur in this experiment.  The leading-order in $\epsilon$
versions of equations (\ref{dCQsA}) establish that (to leading order)
$\mathcal{M}$ decays exponentially.   Using an exponential best fit
empirically determines the only free parameter in the system,
$\delta=0.001107$cm$^{-1}$.

The dissipative theories do a much better job predicting the evolution
of $\mathcal{M}$ and $\mathcal{Q}$ than do the conservative theories.
The Dysthe equation predicts large increases in $\mathcal{P}$ and
$k_m$ that were not observed in the experiment.  It is not possible to
determine which of the other models provides the best prediction for
these quantities because the changes are within the bounds of
experimental measurement error.  However, we note that the vB system
predicts a small FD and hypothesize that FD may have been observed
had the experimental tank been longer.

\subsection{Large-amplitude experiment}

\begin{figure}
\begin{center}
\includegraphics[width=\textwidth]{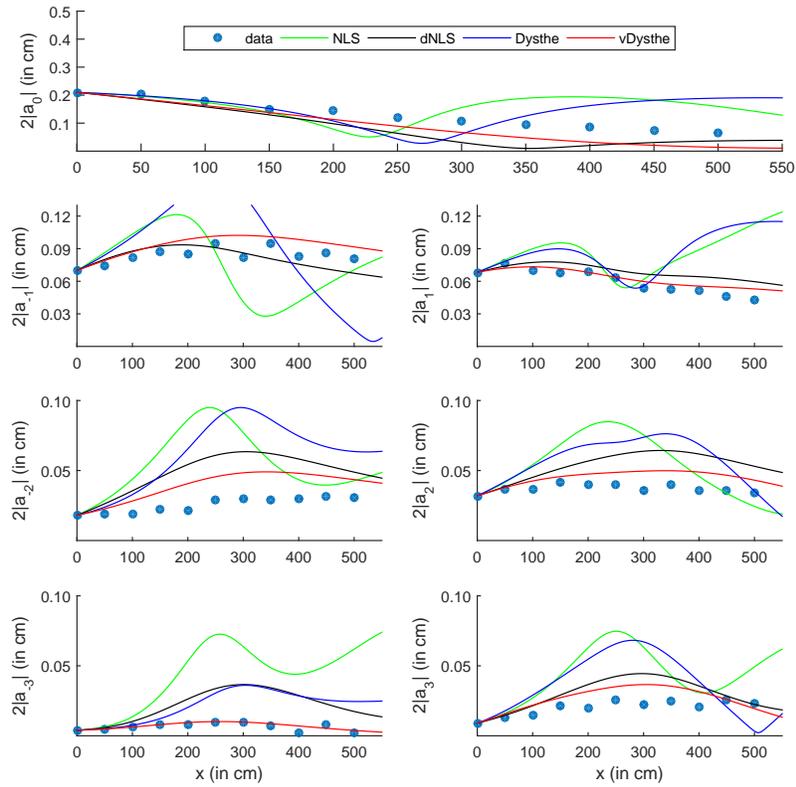}
\caption{Plots of the amplitudes (in cm) of the carrier wave and the
  six nearest sidebands plotted versus the distance (in cm) from
  the first wave gauge for the large-amplitude experiment.}
\label{FampsLar}
\end{center}
\end{figure}

Table \ref{LATable} contains the values of the physical parameters for
this experiment.  Figure \ref{FampsLar} contains plots of the Fourier
amplitudes for this experiment.  In this experiment, the amplitude of
the first lower sideband overtakes that of the carrier wave somewhere
between 300 and 350cm from the first wave gauge.  Therefore FD in the
spectral peak sense occurred in this experiment.

The vB system provides the most accurate prediction for the evolution
for all amplitudes.  It does a good job predicting the evolution of
all seven amplitudes.  The vB system predicts that the amplitude of
the first lower sideband will overtake that of the carrier wave
approximately 223cm from the first wave gauge.  This is earlier than
what was observed in the experiment and is related to the fact that
the vB system slightly over predicts the amplitude of the first lower
sideband.  Regardless, the vB system qualitatively predicts the
evolution of the wave train.  The vA system provides predictions
(results not shown) that are noticeably less accurate than those of
the vB system.  We note that the accuracy of the vB system comes from
a combination of the higher-order viscous term and the higher-order
nonlinear terms.  Predictions from the NLS equation with the two
viscous terms added are much less accurate than those from the vB
system.

\begin{figure}
\begin{center}
\includegraphics[width=\textwidth]{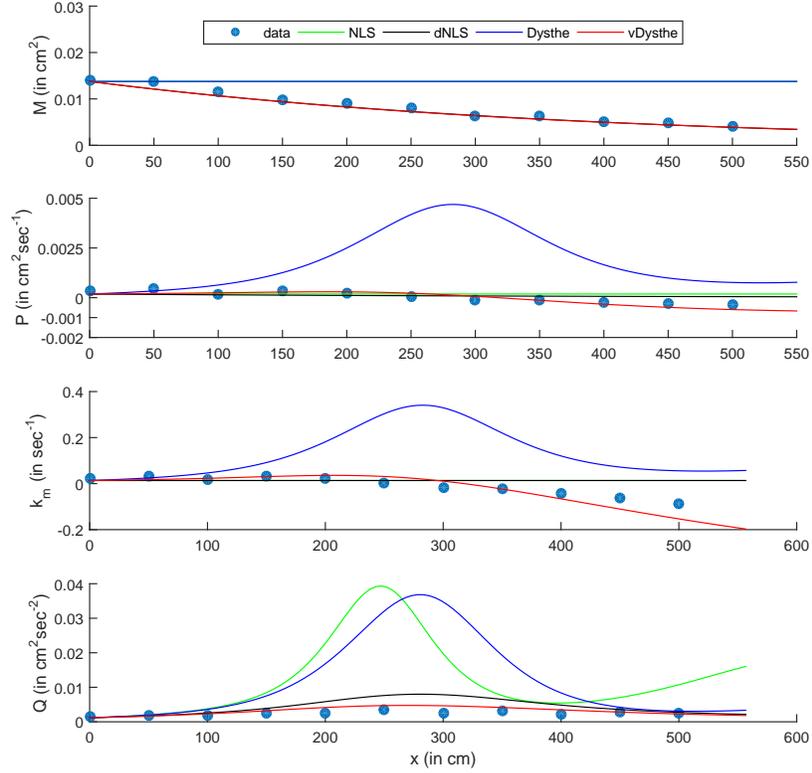}
\caption{Plots of $\mathcal{M}$, $\mathcal{P}$,
  $k_m$, and $\mathcal{Q}$ versus the
  distance (in cm) from the first wave gauge for the large-amplitude
  experiment.}
\label{CQLar}
\end{center}
\end{figure}

Figure \ref{CQLar} shows how the quantities $\mathcal{M}$,
$\mathcal{P}$, $k_m$, and $\mathcal{Q}$ evolved as the wave train
progressed down the tank for the large-amplitude experiment.  The
spectral mean, $k_m$, decreases substantially and therefore FD in the
spectral mean sense occurred in this experiment.  An exponential fit
of the experimental $\mathcal{M}$ data establishes that
$\delta=0.0012697$cm$^{-1}$.

\begin{table}
\begin{center}
\begin{tabular}{|c|c|c|}
\hline
parameter & symbol & value\\
\hline
wave number of carrier wave & $k_0$ & 0.447475\\
initial amplitude of carrier wave & $a_0$ & -0.057134+0.088094i\\
initial amplitude of first upper sideband & $a_1$ & 0.00092806-0.034102i\\
initial amplitude of second upper sideband & $a_2$ &  0.0069104+0.014365i\\
initial amplitude of third upper sideband & $a_3$ & -0.0036015-0.0027146i\\
initial amplitude of first lower sideband & $a_{-1}$ &  0.010223+0.033383i\\
initial amplitude of second lower sideband & $a_{-2}$ &  0.0088658+0.0016430i\\
initial amplitude of third lower sideband & $a_{-3}$ &  -0.0014867-0.0013086i\\
dimensionless parameter & $\epsilon=2k_0|a_0|$ & 0.093970\\
viscosity coefficient & $\delta$ & 0.0012697\\
\hline
\end{tabular}
\caption{Experimental parameters for the large-amplitude experiment.}
\label{LATable}
\end{center}
\end{table}

The vB system provides the best predictions for all four of these
quantities.  It predicts a brief temporary frequency upshift
followed by a downward trend.  The Dysthe equation predicts a
temporary, but dramatic, frequency upshift that is quite different than
what was experimentally observed.  Both the NLS and dissipative NLS
predict that $k_m$ is constant even though the experimental data shows
a definite downward trend.  Note that the plots of $\mathcal{P}$ and
$k_m$ make it obvious that the initial conditions used in the
numerical simulations of the PDEs do not exactly match the
experimental initial conditions.  This is because only the carrier
wave and the first six sidebands are used in the initial conditions
(see equation (\ref{ICs})).  If more modes are included, the initial
points line up more closely, but the qualitative behavior of the plots
remains the same.

We are grateful to Shusen Ding, Diane Henderson, and Harvey Segur for
helpful discussions.  This material is based upon work supported by
the National Science Foundation under grant DMS-1107476.

\end{document}